\documentclass[aip,graphicx]{revtex4-1}
\usepackage{graphicx}
\draft 

\begin{document}


\title{Remotely sensed  transport in microwave photoexcited GaAs/AlGaAs two-dimensional electron system} 



\author{Tianyu Ye}
\author{R. G. Mani}
\affiliation{Department of Physics and Astronomy, Georgia State University, Atlanta, Georgia 30303, USA}

\author{W. Wegscheider}
\affiliation{Laboratorium f\"ur Festk\"orperphysik, ETH Z\"urich, 8093 Z\"urich, Switzerland}

\date{\today}

\begin{abstract}
We demonstrate a strong correlation between the magnetoresistive and the concurrent
microwave reflection from the microwave photo-excited GaAs/AlGaAs  two-dimensional electron system (2DES).  These correlations are
followed as a function of the microwave power, the microwave
frequency, and the applied current.  Notably, the character of the reflection signal remains unchanged even when the current
is switched off in the GaAs/AlGaAs Hall bar specimen. The results
suggest a perceptible microwave-induced change in the electronic
properties of the 2DES, even in the absence of an applied current.
\end{abstract}


\maketitle 

Microwave semiconductor devices have had a profound impact on the fields of satellite and ground based communications, radar, and missile guidance,\cite{bookMicrowave} and related technological developments have fueled motivation for additional basic research on semiconductors such as, for example, the GaAs/AlGaAs two-dimensional electron system. Meanwhile, the ever-improving electron mobility in the high mobility GaAs/AlGaAs 2DES continues to reveal electronic effects induced by microwave and terahertz photo-excitation, such as, for example, the zero-resistance state, without concurrent Hall resistance quantization,\cite{Maninature2002,ZudovPRLDissipationless2003}  observed in the GaAs/AlGaAs 2DES,  when the specimen is
subjected to microwave and terahertz photo-excitation in a magnetic field. The experimental realization of such
radiation-induced zero-resistance states, and associated
$B^{-1}$-periodic radiation-induced magnetoresistance oscillations
expanded the experimental\cite{Maninature2002,ZudovPRLDissipationless2003,ManiPRBVI2004,  ManiPRLPhaseshift2004, ManiEP2DS152004,  KovalevSolidSCommNod2004, SimovicPRBDensity2005,studenikinPRBRef+Abs+Temp2005,ManiPRBTilteB2005, SmetPRLCircularPolar2005,WirthmannPRB2007, WiedmannPRBInterference2008, DennisKonoPRLConductanceOsc2009,ManiPRBPhaseStudy2009,ManiPRBAmplitude2010,FedorychPRBmagnetoabsorption2010, ArunaPRBeHeating2011,ManiPRBPolarization2011,ManinatureComm2012} and
theoretical\cite{DurstPRLDisplacement2003, AndreevPRLZeroDC2003,RyzhiiJPCMNonlinear2003, KoulakovPRBNonpara2003,LeiPRLBalanceF2003, DmitrievPRBMIMO2005, LeiPRBAbsorption+heating2005, InarreaPRLeJump2005, ChepelianskiiEPJB2007,FinklerHalperinPRB2009, DmitrievPRBMixdisorder2009,ChepelianskiiPRBedgetrans2009,HagenmullerPRBCoupling2010, InarreaNanotech, InarreaPRBPower2011, MikhailovPRBponderomotive2011,LindnerNatPhysTI2011,GuPRLIrradiatGraphene2011, LeiPRBMultiPho2011} investigations of transport in the
photo-excited 2-dimensional electron system.

Various mechanisms exist for understanding the radiation-induced
magnetoresistance oscillations including radiation-assisted
indirect inter-Landau-level scattering by phonons and impurities
(the displacement model),\cite{DurstPRLDisplacement2003,
RyzhiiJPCMNonlinear2003, LeiPRLBalanceF2003,
DmitrievPRBMixdisorder2009} non-parabolicity effects in an
ac-driven system (the non-parabolicity
model),\cite{KoulakovPRBNonpara2003} a radiation-induced steady
state non-equilibrium distribution (the inelastic
model),\cite{DmitrievPRBMIMO2005} and the periodic motion of the
electron orbit centers under irradiation (the radiation driven
electron orbit model).\cite{InarreaPRLeJump2005}  According to theory, the experimentally observed zero-resistance states result from either a current instability\cite{AndreevPRLZeroDC2003, FinklerHalperinPRB2009} or from the accumulation/depletion of carriers at the contacts.\cite{MikhailovPRBponderomotive2011}.
\begin{figure}
\centering
\includegraphics[width = 85mm]{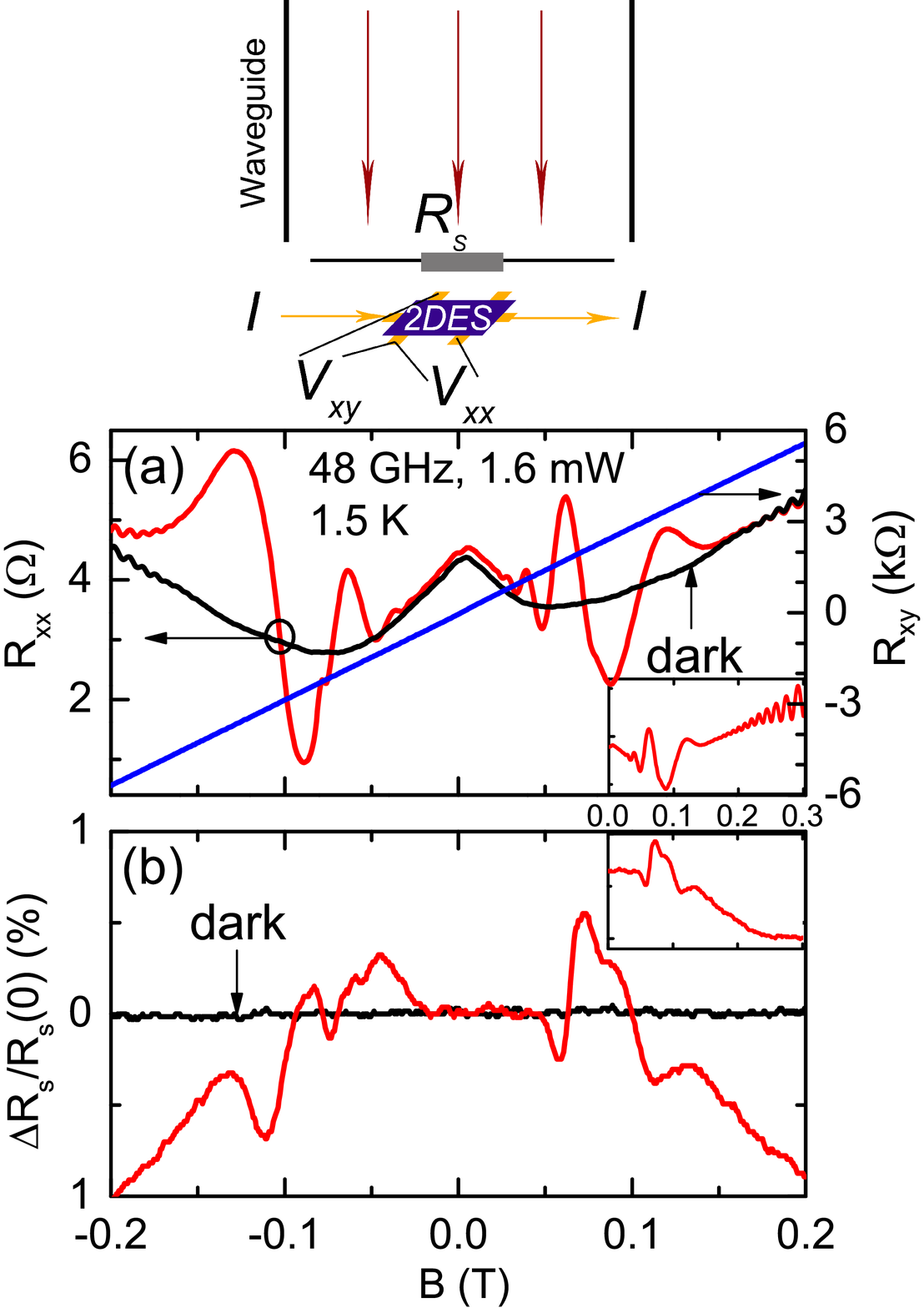}
\caption{(Color online) Top: A schematic of the measurement configuration showing the GaAs/AlGaAs Hall bar and the remote sensing resistor, $R_{s}$, located at the bottom of a cylindrical waveguide, within a low temperature cryostat.
The panels (a) and (b) show the diagonal resistance ($R_{xx}$), Hall
resistance ($R_{xy}$) and the fractional change of the remote detector
resistance ($\Delta$$R_s/R_s(0)$) as functions of magnetic field, $B$, of sample S1.
(a) $R_{xx}$ (left panel) and $R_{xy}$ (right panel)
of S1 with (red curve) and without (black curve) 48 GHz microwave
illumination. (b) Concurrent measurement of $\Delta R_s/R_s(0)$
with (red curve) and without (black curve) 48 GHz microwave
excitation. The insets of (a) and (b) show the photoexcited $R_{xx}$ and $\Delta$$R_s/R_s(0)$ signals over a broader $B$-range.}
\end{figure}

Here,  we compare the oscillatory magnetoresistive response of the microwave
photo-excited GaAs/AlGaAs 2D electron system with the concurrent
microwave reflection that is detected by a nearby resistance
sensor. We report strong correlations, which are
followed as a function of the microwave power, the microwave
frequency, and the applied current. Notably, the character of the reflection
signal remains unchanged even when the current is switched off in
the GaAs/AlGaAs Hall bar specimen. The results suggest a
perceptible microwave-induced change in the electronic properties
of the 2DES, even in the absence of an applied current.



GaAs/AlGaAs Hall bars were mounted at the end of a
long cylindrical waveguide along with a carbon resistor to sense the microwave reflection as shown in Fig. 1. The carbon sensor exhibits a strong negative temperature coefficient, i.e., $dR_{s}/dT \le 0$. Thus, sensor heating by microwaves reflected from the 2DES produces a reduction in the sensor resistance $R_{s}$, which becomes the signature of microwave reflection (or emission).   The waveguide sample holder was
inserted into a variable temperature insert in a superconducting solenoid. A base temperature of approximately
$1.5$ K was realized by pumping on the liquid helium within the
variable temperature insert. The 2D electron
density $n \approx 2.3 \times 10^{11}$ cm$^{-2}$ and the mobility $\mu \approx 8 \times 10^{6}$ cm$^{2}$/Vs. Thus, the transport lifetime is $\tau \approx 3 \times 10^{-10}$ s and the
single particle lifetime $\tau_{s} \approx 2.8 \times 10^{-12}$ s.
Finally, a low frequency lock-in technique was
adopted to detect the electrical signals of interest.

\begin{figure}
\centering
\includegraphics[width = 85mm]{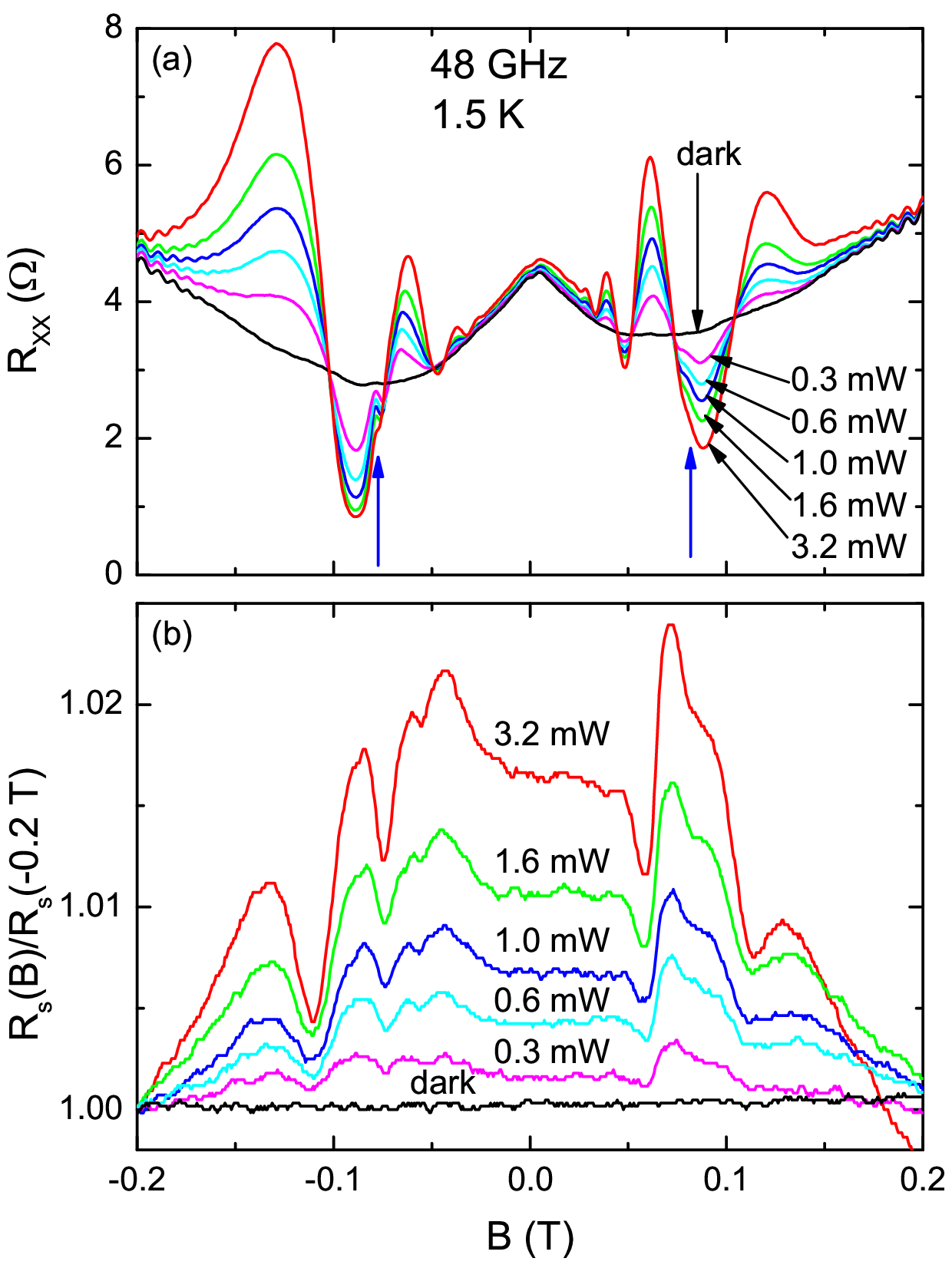}
\caption{(Color online) (a) The dark- and photoexcited- at 48 GHz
$R_{xx}$ signal,  and (b) the concurrently measured normalized
remotely sensed signal $R_s/R_{s}(-0.2$ T$)$ for a high mobility
GaAs/AlGaAs specimen. Various colored traces correspond to the
different power levels over the range  $ 0 \le p \le 3.2$ mW. The
blue upward arrows in (a) point out the inflections on the
oscillatory resistance.}
\end{figure}

Figure 1 compares the diagonal resistance $R_{xx}$ (Fig. 1(a)) of a
high mobility GaAs/AlGaAs sample and the concurrently measured
remote detector resistance $R_s$ (Fig. 1(b)). Fig. 1(a)
shows that the sample in the dark, i.e., without microwave
photo-excitation, does not exhibit microwave-radiation-induced
magneto-resistance oscillations in $R_{xx}$, as the fractional
change in the reflection detector signal $\Delta R_{s}/R_{s}$, see
Fig. 1(b), is featureless as well in the dark. However, with $48$
GHz microwave photoexcitation, see Fig. 1(a) and Fig. 1(b), $R_{xx}$ exhibits microwave
induced magnetoresistance oscillations for $-0.2 \le B \le 0.2$
Tesla, see the red trace in Fig. 1(a), and $\Delta R_{s}/R_{s}$
conveys an oscillatory reflection. Here,
the microwave-induced changes in the detector signal
$\Delta$$R_s/R_s(0)\approx$$1\%$, where $\Delta$$R_s$=$R_s(B)-R_s(0)$, $R_{s}(B)$ is the sensor resistance in the magnetic field, and $R_{s} (0)$ is the zero-field sensor resistance.  To confirm that the microwave induced effect in $\Delta R_{s}/R_{s}$ is due to the microwave response of the high-mobility GaAs/AlGaAs specimen, we have also examined low mobility GaAs/AlGaAs specimens in the same experimental setup. The results showed little difference between the photo-excited- and dark- $R_{xx}$, and also $\Delta R_{s}/R_{s}$ in the low mobility specimen, which is attributed to the absence of radiation-induced oscillations in the low mobility specimen. Finally,  the insets of Fig. 1 (a) and (b) show that where $R_{xx}$ in the high mobility specimen exhibits only Shubnikov-de Haas oscillations, $R_s$ monotonically decreases as $B$ is
 ncreased, but oscillatory features are not observable. 

\begin{figure}
\centering
\includegraphics[width = 85mm]{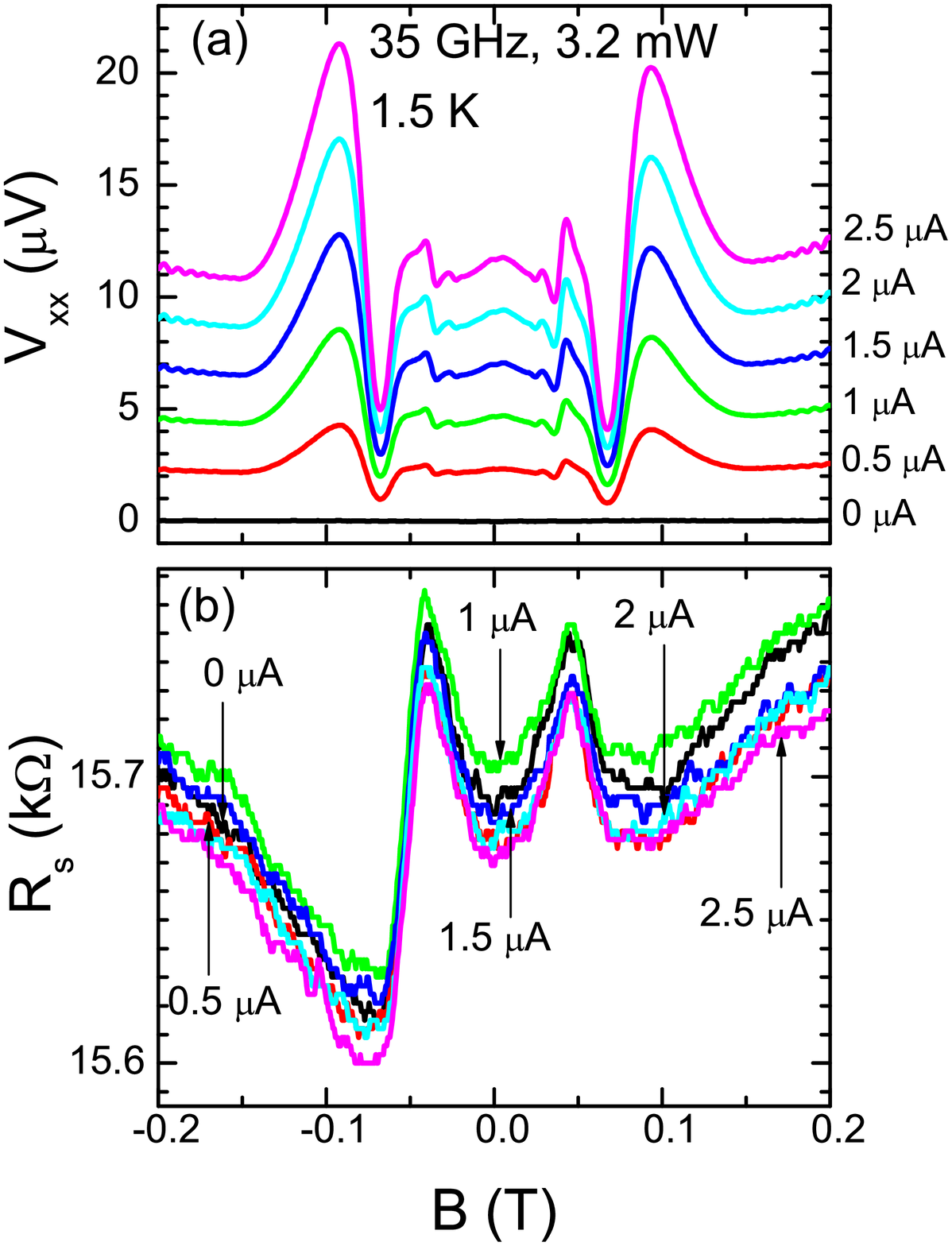}
\caption{(Color online) The diagonal voltage $V_{xx}$ and the
remote sensor resistance $R_s$ are exhibited for a GaAs/AlGaAs
specimen S1 under 35 GHz microwave excitation. The different color
curves correspond to discrete applied currents, I, through sample
with $0 \le I \le 2.5$ $\mu$A. The same color code has been used
in the top and bottom panels. }
\end{figure}

Figure 2 exhibits the microwave power ($P$)-dependence of $R_{xx}$- and the normalized $R_{s}$-, i.e., $R_{s}/R_{s}(-0.2$ T$)$, vs. $B$. The exhibited traces indicate the following
features with changing microwave power: i) The phase of the
microwave radiation-induced magnetoresistance oscillations in
$R_{xx}$ does not change with the microwave power, but oscillation
amplitude increases with the power, $P$, over the range $0.3\le P
\le$ 3.2 mW. ii)Correspondingly,  the oscillatory reflection conveyed by 
$R_s/R_{s}(-0.2$ T$)$ remains in phase for different microwave
powers, although the relative amplitude increases with  $P$.  Note that the reflection response in Fig. 2  is not just confined to the vicinity of cyclotron resonance ($B \approx 0.11$ Tesla).

\begin{figure*}
\centering
\includegraphics[width = 170mm]{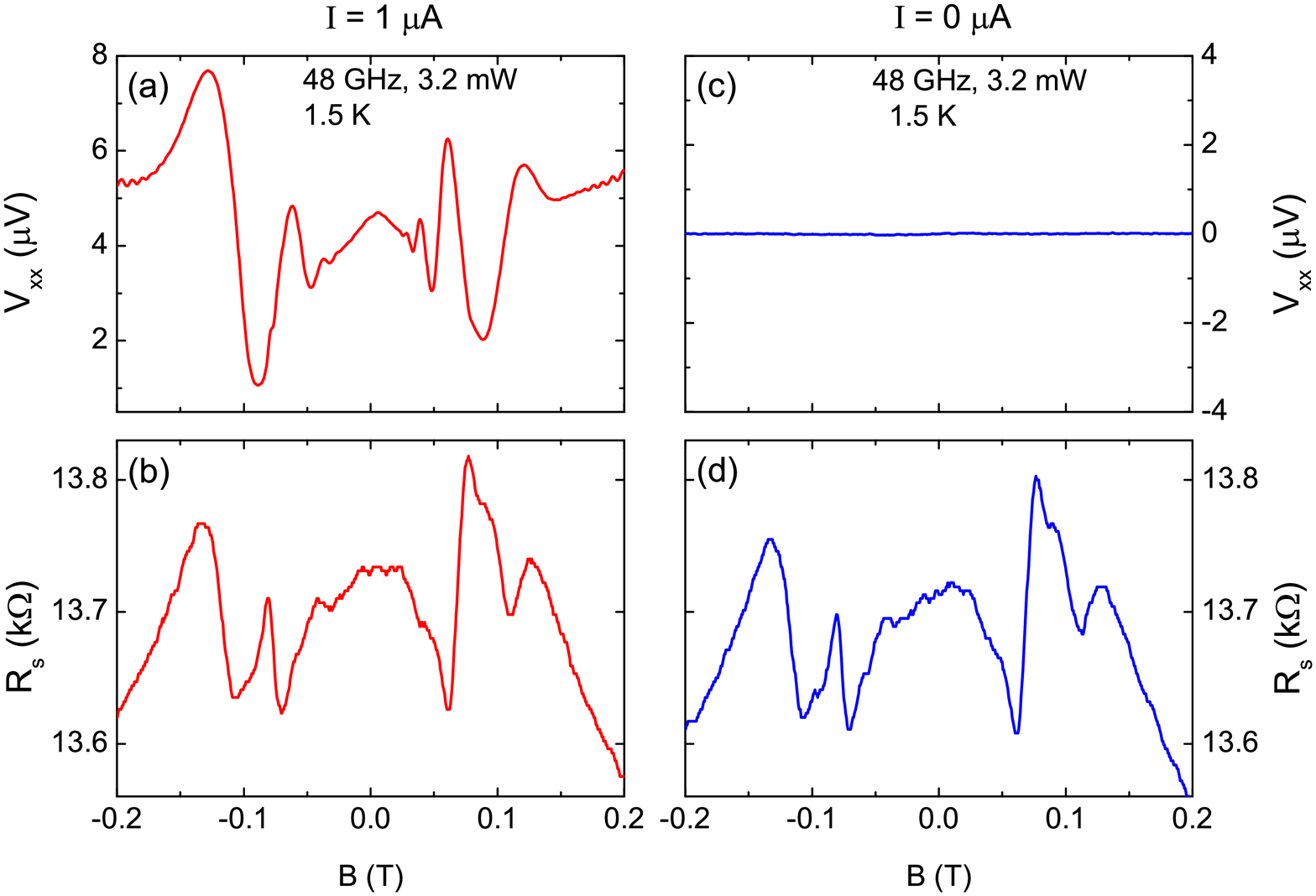}
\caption{(Color online)The diagonal voltage  $V_{xx}$ and the
remotely sensed signal $R_s$ for a GaAs/AlGaAs 2DES subjected to
48 GHz microwave excitation. In plots (a) and (b), the applied
current $ I = 1$ $\mu$A. Plots (c) and (d) correspond to $I = 0$
$\mu$A. Note that the $R_{s}$ signal remains unchanged upon
switching off the current through the specimen.}
\end{figure*}

Figure 3 examines the current ($I$) dependence of $V_{xx} = IR_{xx}$ and  $R_{s}$ at 35 GHz and $P = 3.2$ mW. Here $V_{xx}$ has been plotted rather than $R_{xx}$ because $R_{xx}$ is undefined when $I = 0$ $\mu$A. As the applied
current, $I$, is decreased from 2.5 $\mu$A to 0 $\mu$A,
the diagonal voltage $V_{xx}$ decreases proportionally, see Fig.
3(a), as expected.  Fig. 3(b) exhibits the
concurrently measured remotely sensed signal $R_{s}$ at the same
currents, $I$. Here, note the insensitivity in the $R_{s}$ signal to the applied current. 

To convey the remarkable feature in this result, in Fig. 4, we
exhibit the $V_{xx}$ vs. $B$ and $R_s$ vs. $B$ now with 48 GHz
microwave excitation at 3.2 mW, with the applied current $I = 1$
$\mu$A, in Fig. 4(a) and Fig. 4(b), and $I= 0$ $\mu$A in Fig. 4(c)
and Fig. 4(d). When
the applied current is switched off, the $V_{xx}$ signal vanishes
as illustrated in Fig. 4(c). However, the non-monotonic variation
in $R_{s}$ due to the microwave reflection from the photoexcited 2DES persists even in the absence of the applied current. These features indicate that there is a
microwave-induced response in the 2DES even in the absence of an
applied current.


We provide a brief explanation of the observed features using the theory of Lei and Liu.\cite{LeiPRLBalanceF2003, LeiPRBAbsorption+heating2005} They asserted that,
in the presence of impurity and phonon scattering, which couple
the c.m. and relative motions, the microwave field affects the
relative motion by allowing transitions between different states,
leading to radiation-induced magnetoresistance
oscillations.\cite{LeiPRLBalanceF2003} Their calculations indicated oscillations in the energy absorption rate, $S_{p}$, which
correlated with resonant oscillations in the electron
temperature.\cite{LeiPRBAbsorption+heating2005} Although the theory did not
explicitly examine the $S_{p}$ in the limit of a vanishing applied
electric field or current, it appears plausible that the energy
absorption rate of the 2DES might be independent of whether or not
an applied electric field or current exists in the specimen. If
the $S_{p}$ continued to exhibit resonant oscillations even
without an applied current or electric field, then it seems
plausible that the reflected microwave power would follow the
oscillatory $S_{p}$ and also exhibit oscillations as seen in the
experimental data here. Therefore, one might tentatively attribute the oscillations
in the remotely sensed signal $R_{s}$ to the oscillatory variation
in the energy absorption rate of the microwave photo-excited 2DES
and the concomitant change in the reflection.

In the radiation driven electron orbit model,\cite{InarreaPRLeJump2005}
one expects a periodic back- and forth- radiation driven
motion of the electron orbits. Since such oscillatory motion
of electron charge is, from the classical perspective,
expected to produce radiation, such reflection/emission
signal reported by $R_{s}$ might be expected.
A full theory of this has not yet been published.


In summary, the magnetoresistive response of the microwave photo-excited
GaAs/AlGaAs 2D electron system has been compared with the
concurrent microwave reflection from the 2DES. The experimental
results indicate a strong correlation between the observed
features in the two types of measurements.  Curiously, the character of the
reflection signal remains unchanged even when the current is
switched off in the GaAs/AlGaAs Hall bar specimen. The results
suggest that the 2DES is microwave active even in the absence of
an applied current. 


Basic research at Georgia State University (GSU) is  supported
by the DOE-BES, MSE Division under DE-SC0001762. Additional
support for microwave work is provided by the ARO under
W911NF-07-01-0158.

%
%

%



\pagebreak
\section{Figure Captions}
Figure 1: (Color online) Top: A schematic of the measurement
configuration showing the GaAs/AlGaAs Hall bar and the remote
sensing resistor, $R_{s}$, located at the bottom of a cylindrical
waveguide, within a low temperature cryostat. The panels (a) and
(b) show the diagonal resistance ($R_{xx}$), Hall resistance
($R_{xy}$) and the fractional change of the remote detector
resistance ($\Delta$$R_s/R_s(0)$) as functions of magnetic field,
$B$, for sample S1. (a) $R_{xx}$ (left panel) and $R_{xy}$ (right
panel) of S1 with (red curve) and without (black curve) 48 GHz
microwave illumination. (b) Concurrent measurement of $\Delta
R_s/R_s(0)$ with (red curve) and without (black curve) 48 GHz
microwave excitation. The insets of (a) and (b) show the
photoexcited $R_{xx}$ and $\Delta$$R_s/R_s(0)$ signals over a
broader $B$-range.

Figure 2: (Color online) (a) The dark- and photoexcited- at 48 GHz
$R_{xx}$ signal,  and (b) the concurrently measured normalized
remotely sensed signal $R_s/R_{s}(-0.2$ T$)$ for a high mobility
GaAs/AlGaAs specimen. Various colored traces correspond to the
different power levels over the range  $ 0 \le p \le 3.2$ mW. The
blue upward arrows in (a) point out the inflections on the
oscillatory resistance.

Figure 3: (Color online) The diagonal voltage $V_{xx}$ and the
remote sensor resistance $R_s$ are exhibited for a GaAs/AlGaAs
specimen S1 under 35 GHz microwave excitation. The different color
curves correspond to discrete applied currents, I, through sample
with $0 \le I \le 2.5$ $\mu$A. The same color code has been used
in the top and bottom panels.

Figure 4: (Color online)The diagonal voltage  $V_{xx}$ and the
remotely sensed signal $R_s$ for a GaAs/AlGaAs 2DES subjected to
48 GHz microwave excitation. In plots (a) and (b), the applied
current $ I = 1$ $\mu$A. Plots (c) and (d) correspond to $I = 0$
$\mu$A. Note that the $R_{s}$ signal remains unchanged upon
switching off the current through the specimen.

\pagebreak
\begin{figure}[t]
\centering
\includegraphics[width = 85mm]{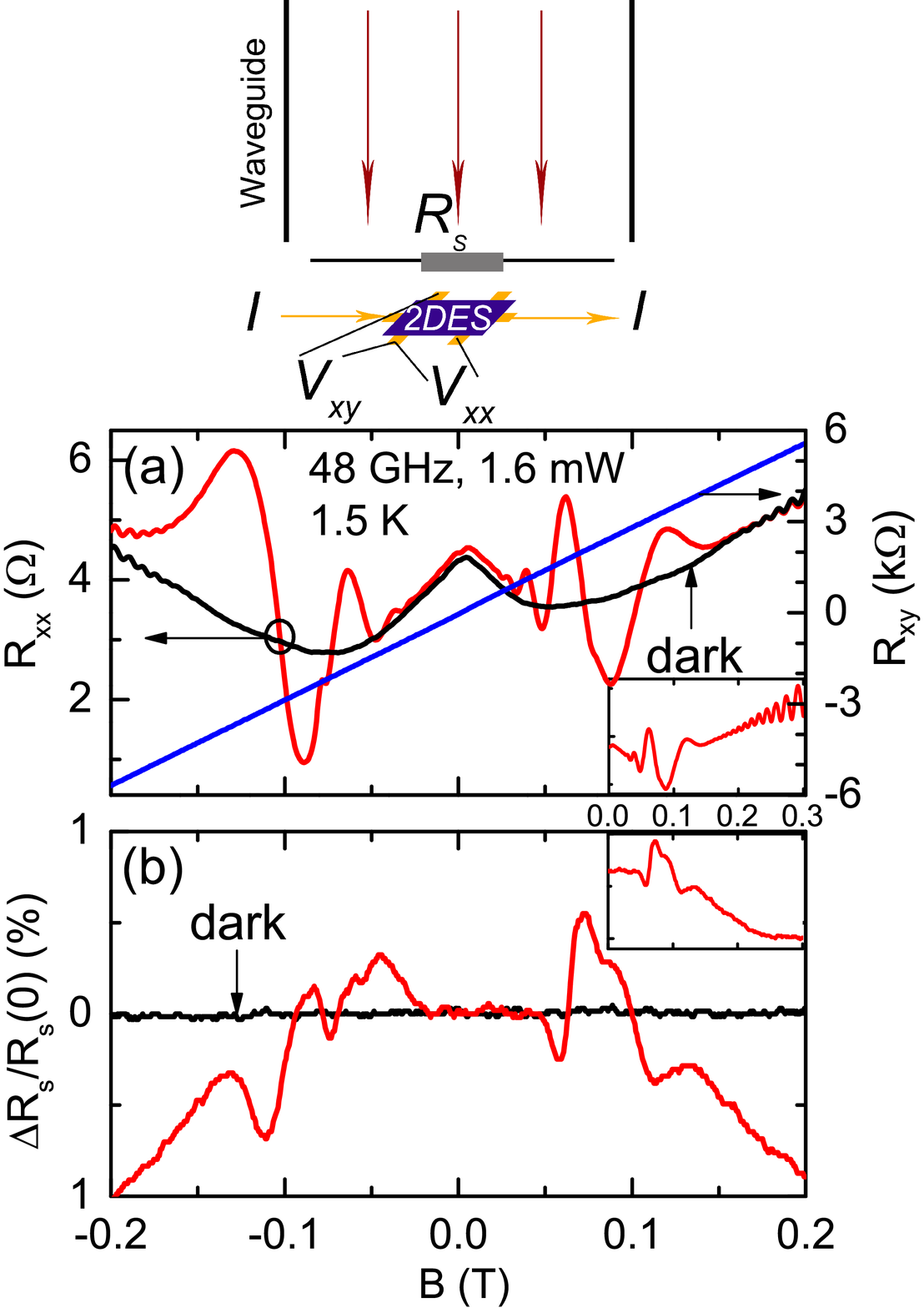}
\begin{center}
Figure 1
\end{center}
\end{figure}

\begin{figure}[t]
\centering
\includegraphics[width = 85mm]{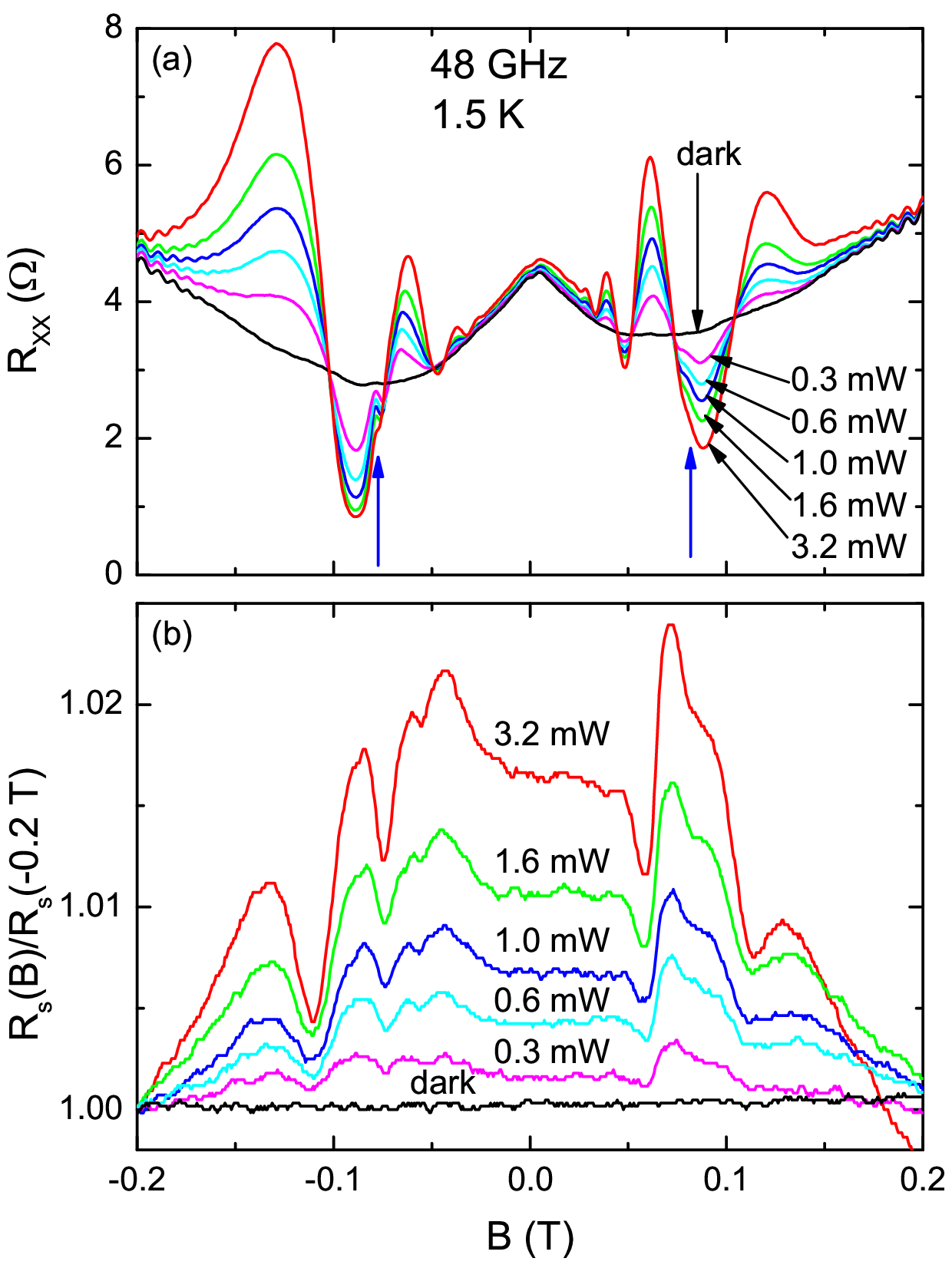}
\begin{center}
Figure 2
\end{center}
\end{figure}

\begin{figure}[t]
\centering
\includegraphics[width = 85mm]{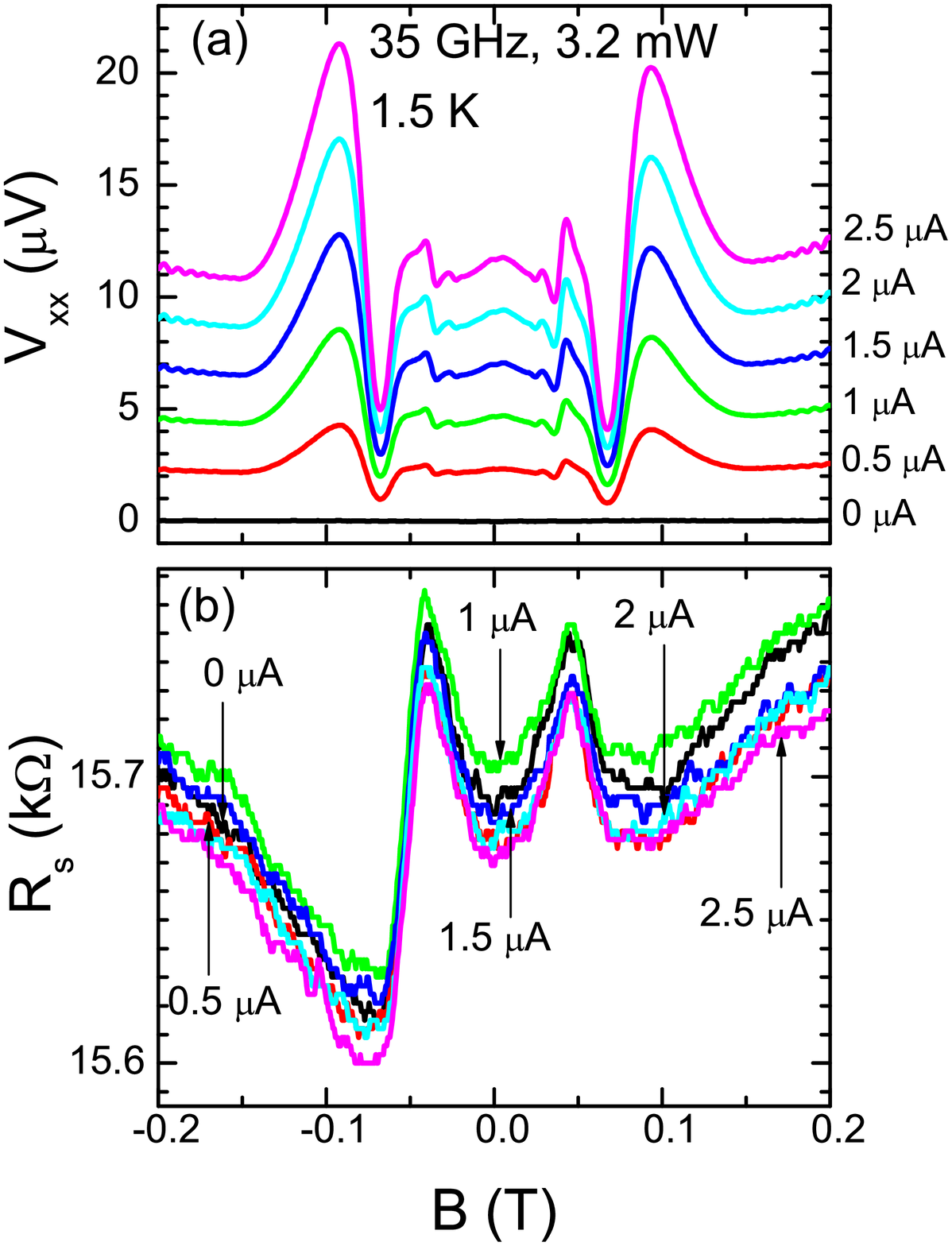}
\begin{center}
Figure 3
\end{center}
\end{figure}

\begin{figure*}[t]
\centering
\includegraphics[width = 170mm]{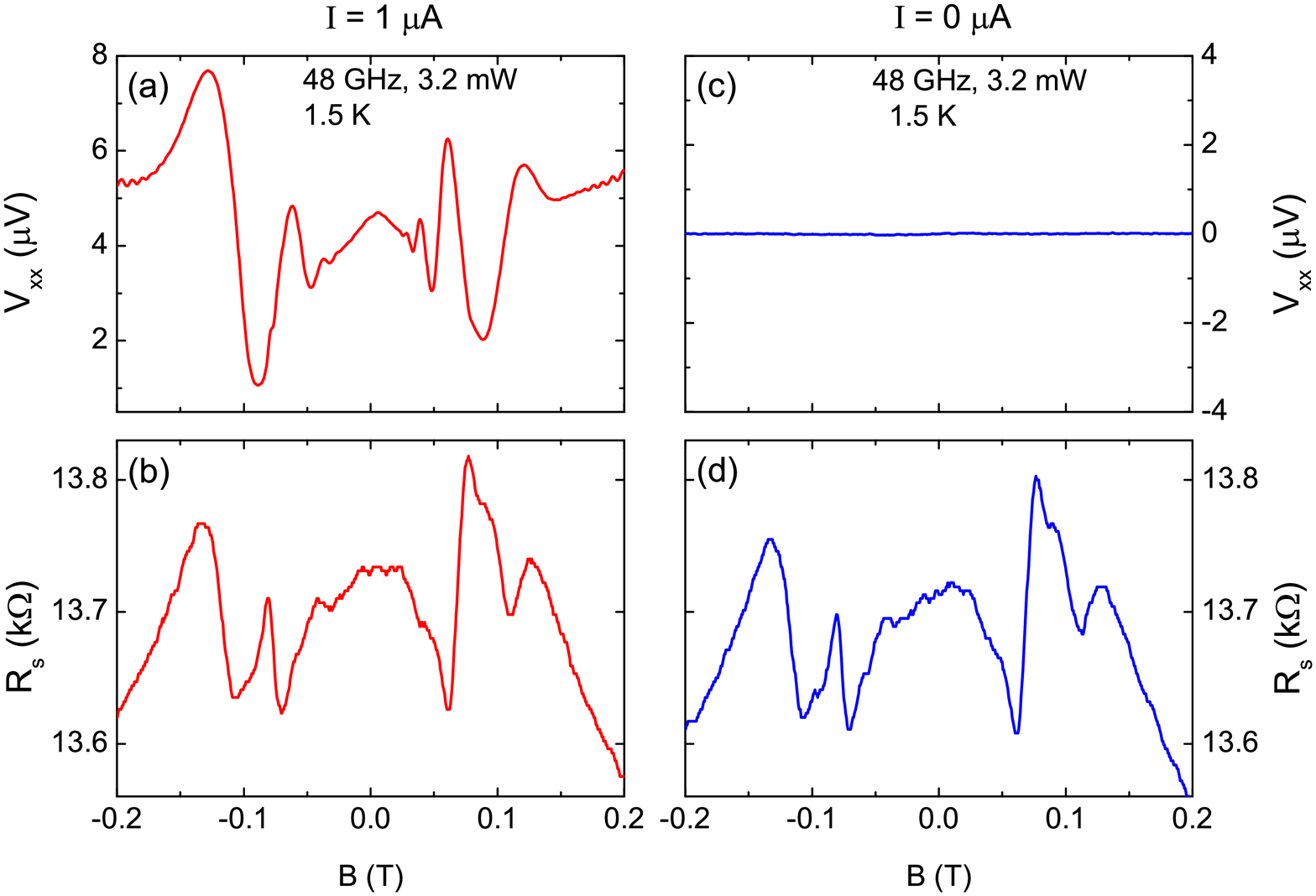}
\begin{center}
Figure 4
\end{center}
\end{figure*}


\begin{thebibliography}{0}

\bibitem{bookMicrowave} J-F. Luy, \textit{Microwave Semiconductor Devices: Theory, Technology, and Performance},(Expert-Verlag, Renningen, 2005).
\bibitem{Maninature2002} R. G. Mani, J. H. Smet, K. von Klitzing, V. Narayanamurti, W. B. Johnson, and V. Umansky, Nature \textbf{420}, 646 (2002).
\bibitem{ZudovPRLDissipationless2003} M. A. Zudov, R. R. Du, L. N. Pfeiffer, and K. W. West, Phys. Rev. Lett. \textbf{90}, 046807 (2003).
\bibitem{ManiPRBVI2004} R. G. Mani, V. Narayanamurti, K. von Klitzing, J. H. Smet, W. B. Johnson, and V. Umansky, Phys. Rev. B \textbf{70}, 155310 (2004);  Phys. Rev. B \textbf{69}, 161306 (2004).
\bibitem{ManiPRLPhaseshift2004} R. G. Mani, J. H. Smet, K. von Klitzing, V. Narayanamurti, W. B. Johnson, and V. Umansky, Phys. Rev. Lett. \textbf{92}, 146801 (2004); Phys. Rev. B \textbf{69}, 193304 (2004).

\bibitem{ManiEP2DS152004} R. G. Mani, Physica E \textbf{22}, 1 (2004); Physica E. \textbf{25}, 189 (2004).
\bibitem{KovalevSolidSCommNod2004} A. E. Kovalev, S. A. Zvyagin, C. R. Bowers, J. L. Reno, and J. A. Simmons, Solid State Commun. \textbf{130}, 379 (2004).
\bibitem{SimovicPRBDensity2005} B. Simovi\ifmmode \check{c}\else \v{c}\fi{}, C. Ellenberger, K. Ensslin, H. P. Tranitz, and W. Wegscheider, Phys. Rev. B \textbf{71}, 233303 (2005).
\bibitem{studenikinPRBRef+Abs+Temp2005} S. A. Studenikin, M. Potemski, A. Sachrajda, M. Hilke, L. N. Pfeiffer, and K. W. West, Phys. Rev. B \textbf{71}, 245313 (2005).
\bibitem{ManiPRBTilteB2005} R. G. Mani, Phys. Rev. B \textbf{72}, 075327 (2005); Appl. Phys. Lett. \textbf{91}, 132103 (2007);  Appl. Phys. Lett. \textbf{92}, 102107 (2008); Physica E40, 1178 (2008)
\bibitem{SmetPRLCircularPolar2005} J. H. Smet, B. Gorshunov, C. Jiang, L. Pfeiffer, K. West, V. Umansky, M. Dressel, R. Meisels, F. Kuchar, and K. von Klitzing, Phys. Rev. Lett. \textbf{95}, 116804 (2005).
\bibitem{WirthmannPRB2007} A. Wirthmann, B. D. McCombe, D. Heitmann, S. Holland, K. Friedland, and C. M. Hu, Phys. Rev. B \textbf{76}, 195315 (2007).
\bibitem{WiedmannPRBInterference2008} S. Wiedmann, G. M. Gusev, O. E. Raichev, T. E. Lamas, A. K. Bakarov, and J. C. Portal, Phys. Rev. B \textbf{78}, 121301 (2008).
\bibitem{DennisKonoPRLConductanceOsc2009} D. Konstantinov and K. Kono, Phys. Rev. Lett. \textbf{103}, 266808 (2009).
\bibitem{ManiPRBPhaseStudy2009} R. G. Mani, W. B. Johnson, V. Umansky, V. Narayanamurti, and K. Ploog, Phys. Rev. B \textbf{79}, 205320 (2009).
\bibitem{ManiPRBAmplitude2010} R. G. Mani, C. Gerl, S. Schmult, W. Wegscheider, and V. Umansky, Phys. Rev. B \textbf{81}, 125320 (2010).
\bibitem{FedorychPRBmagnetoabsorption2010} O. M. Fedorych, M. Potemski, S. A. Studenikin, J. A. Gupta, Z. R. Wasilewski, and I. A. Dmitriev, Phys. Rev. B \textbf{81}, 201302 (2010).
\bibitem{ArunaPRBeHeating2011} A. N. Ramanayaka, R. G. Mani, and W. Wegscheider, Phys. Rev. B \textbf{83}, 165303 (2011).
\bibitem{ManiPRBPolarization2011} R. G. Mani, A. N. Ramanayaka, W. Wegscheider, Phys. Rev. B \textbf{84}, 085308 (2011); A. N. Ramanayaka, R. G. Mani, J. I\~narrea, and W. Wegscheider, Phys. Rev. B \textbf{85}, 205315 (2012)..
\bibitem{ManinatureComm2012} R. G. Mani, J. Hankinson, C. Berger, and W. A. de Heer, Nat. Commun. \textbf{3}, 996 (2012).
\bibitem{DurstPRLDisplacement2003} A. C. Durst, S. Sachdev, N. Read, and S. M. Girvin, Phys. Rev. Lett. \textbf{91}, 086803 (2003).
\bibitem{AndreevPRLZeroDC2003} A. V. Andreev, I. L. Aleiner, and A. J. Millis, Phys. Rev. Lett. \textbf{91}, 056803 (2003).
\bibitem{RyzhiiJPCMNonlinear2003} V. Ryzhii and R. Suris, J. Phys. Condens. Matter \textbf{15}, 6855 (2003).
\bibitem{KoulakovPRBNonpara2003} A. A. Koulakov and M. E. Raikh, Phys. Rev. B \textbf{68}, 115324 (2003).
\bibitem{LeiPRLBalanceF2003} X. L. Lei and S. Y. Liu, Phys. Rev. Lett. \textbf{91}, 226805 (2003).
\bibitem{DmitrievPRBMIMO2005} I. A. Dmitriev, M. G. Vavilov, I. L. Aleiner, A. D. Mirlin, and D. G. Polyakov, Phys. Rev. B \textbf{71}, 115316 (2005).
\bibitem{LeiPRBAbsorption+heating2005}  X. L. Lei and S. Y. Liu, Phys. Rev. B \textbf{72}, 075345 (2005).
\bibitem{InarreaPRLeJump2005} J. I\~narrea and G. Platero, Phys. Rev. Lett. \textbf{94}, 016806 (2005);  Phys. Rev. B \textbf{76}, 073311 (2007).
\bibitem{ChepelianskiiEPJB2007} A. D. Chepelianskii, A. S. Pikovsky, and D. L. Shepelyansky, Eur. Phys. J. B \textbf{60}, 225 (2007).
\bibitem{FinklerHalperinPRB2009} I. G. Finkler and B. I. Halperin, Phys. Rev. B \textbf{79}, 085315 (2009).
\bibitem{DmitrievPRBMixdisorder2009} I. A. Dmitriev, M. Khodas, A. D. Mirlin, D. G. Polyakov, and M. G. Vavilov, Phys. Rev. B \textbf{80}, 165327 (2009).
\bibitem{ChepelianskiiPRBedgetrans2009} A. D. Chepelianskii, and D. L. Shepelyansky, Phys. Rev. B \textbf{80}, 241308 (2009).
\bibitem{HagenmullerPRBCoupling2010} D. Hagenm\"uller,  S. De Liberato, and C. Ciuti, Phys. Rev. B \textbf{81}, 235303 (2010).
\bibitem{InarreaNanotech} J. Inarrea and G. Platero, Nanotechnology \textbf{21}, 315401 (2010).
\bibitem{InarreaPRBPower2011} J. I\~narrea, R. G. Mani, and W. Wegscheider, Phys. Rev. B \textbf{82}, 205321 (2010).
\bibitem{MikhailovPRBponderomotive2011} S. A. Mikhailov, Phys. Rev. B \textbf{83}, 155303 (2011).
\bibitem{LindnerNatPhysTI2011} N. H. Lindner, G. Refael, and V. Galitski, Nat. Phys. \textbf{7}, 490 (2011).
\bibitem{GuPRLIrradiatGraphene2011} Z. Gu, H. A. Fertig, D. P. Arovas, and A. Auerbach, Phys. Rev. Lett. \textbf{107}, 216601 (2011).
\bibitem{LeiPRBMultiPho2011} X. L. Lei and S. Y. Liu, Phys. Rev. B \textbf{84}, 035321 (2011); Phys. Rev. B \textbf{86}, 205303 (2012).
\end{thebibliography}
\end{document}